\documentclass[rnote]{aa} 
\usepackage{graphicx}
\usepackage{txfonts}
\begin{document}
   \title{On the central ionizing star of G23.96+0.15 and near-IR
spectral classification of O stars}
\titlerunning{On the ionizing star of G23.96+0.15}

   \author{Paul A. Crowther
          \and
          James P. Furness\fnmsep\thanks{Based on observations
made with ESO telescopes at the Paranal Observatory under programme
ID 077.C-0550(A)}
          }

   \offprints{P.A. Crowther}

   \institute{Department of Physics \& Astronomy,
              University of Sheffield,               Hounsfield Road, Sheffield,
              S3 7RH, UK\\
              \email{Paul.Crowther@sheffield.ac.uk}
             }

   \date{Received; accepted}

  \abstract
   {}
   {A near-infrared study of the main ionizing star of the ultracompact HII 
region  G23.96+0.15 (IRAS 18317--0757) is presented, along with a 
re-evaluation  of the distance to 
this source, and a re-assessment of H- and K-band classification 
diagnostics for O dwarfs}
   {We have obtained near-IR VLT/ISAAC imaging and spectroscopy of
G23.96+0.15, plus archival imaging from UKIRT/UFTI. A spectroscopic 
analysis was carried out using a non-LTE model atmosphere code.}
   {A quantitative H- and K-band classification scheme for O dwarfs 
is provided, from which we establish an O7.5V spectral subtype for the 
central star  of G23.96+0.15. We estimate an effective temperature of 
$T_{\rm eff}\sim$38\,kK from a spectral analysis.}
   {A spectroscopic distance of 2.5\,kpc is obtained for
G23.96+0.15, substantially lower than the kinematic distance 
of 4.7\,kpc, in common with recent studies of other Milky way H\,{\sc 
ii} regions. Such discrepancies would be alleviated if sources are 
unresolved binaries or clusters.}


   \maketitle
%

\section{Introduction}

The formation of high mass stars remains an unresolved problem
in contemporary astrophysics (Zinnecker \& Yorke 2007). Despite 
considerable theoretical progress in recent years, this topic
remains observationally challenging. O stars form
deeply embedded within their natal cocoons, presumably within
intermediate to high mass star clusters, only to 
visually emerge after $\sim$0.5\,Myr (Prescott et al. 2007), a 
 a substantial fraction of their 2--10\,Myr main-sequence 
lifetimes.

Radio continuum surveys of ultracompact H\,{\sc ii} (UCHII) regions betray the 
presence of O stars through the effect of their Lyman continuum photons to 
free-free (thermal) emission. Typically, these regions can not be observed 
at wavelengths shorter than the mid-infrared, such that the ionizing stars 
can only be studied indirectly through the circumstellar dust and gas 
(e.g. Peeters et al. 2002; Mart\'{i}n-Hern\'{a}ndez et al. 2002). However, in a 
few instances our sight line to the central source is sufficiently clear 
of circumstellar dust that the O star responsible for the H\,{\sc ii} region can 
be directly observed at near-infrared wavelengths (Watson \& Hanson 1997; 
Hanson et al. 2002; Bik et al. 2005).

Over the past decade, the advent of efficient near-IR spectrographs at 
large ground-based telescopes has permitted medium resolution 
($R\sim 5000$) spectroscopy 
of template Milky Way O stars in the H- and K-bands (Hanson et al. 2005b), 
which has been extended to the O4--5 star responsible for the G29.96--0.02 
UCHII region (IRAS 8434--0242, Hanson et al. 2005a). Spectroscopic 
analysis of near-IR hydrogen and helium lines of O stars agrees closely 
with optical diagnostics in most cases (Repolust et al. 2005). 

Thus far, G29.96--0.02 represents the sole example of an UCHII region
whose ionizing star has been analysed based upon medium spectral 
resolution. The significance of such objects is that, in principal, they 
may serve as calibrators for the so-called `inverse problem', in which
the ionizing stars of embedded UCHII regions may be obtained from 
analysis of mid-infrared fine-structure nebular lines (e.g. 
Mart\'{i}n-Hern\'{a}ndez et al. 2002; Morisset et al. 2004). 

Fortunately, other
UCHII regions are also accessible to near-IR spectroscopy. One such
region, G23.96+0.15 (IRAS 18317--0757) is the focus of the present study, 
for  which Hanson et al. (2002) estimated a spectral type of O7--8 from 
low resolution K-band spectroscopy and Kim \& Koo (2001) have 
inferred a mid-O spectral type from radio continuum observations 
for an adopted distance of 6\,kpc.  Hunter et al. (2004) have 
identified a cluster of embedded massive stars from millimetre 
observations of G23.96+0.15. 

In the present study, we present new near-IR imaging and spectroscopy
of G23.96+0.15 permitting the subtype of the ionizing star to be refined
from H-band and K-band hydrogen and helium line ratios, plus a 
spectroscopic distance for comparison with kinematic results.


\section{Observations}

H-band and K-band spectroscopy of G23.96+0.15 was observed with the ISAAC 
instrument (Moorwood et al. 1998) mounted at the Very Large Telescope 
between 22 April-2 May 2006.  High spatial resolution imaging was drawn 
from acquisition ISAAC datasets, which were supplemented by archival UKIRT 
Fast Trace Imager (UFTI, Roche et al. 2002) K-band imaging, plus H-band 
imaging from Hanson et al. (2002) and 2MASS datasets.

\subsection{Near-IR imaging}

Acquisition ISAAC images of G23.96+0.15 were obtained using the 
1024$\times$1024 Hawaii Rockwell array (0.148 arcsec/pix), with a 
combination of both the 2.17$\mu$m and 2.19$\mu$m narrow-band filters 
during excellent seeing conditions of 0.3--0.5 arcsec in April-May 
2006. 

\begin{figure}
   \centering
   \includegraphics[width=0.45\columnwidth]{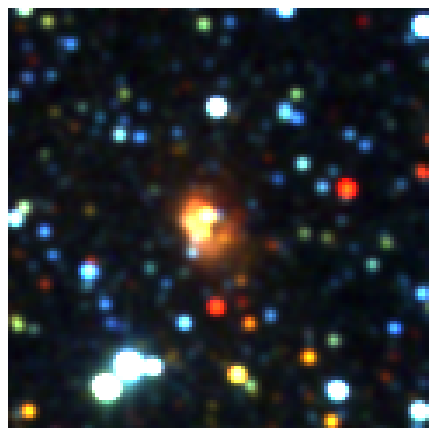}
   \includegraphics[width=0.51\columnwidth]{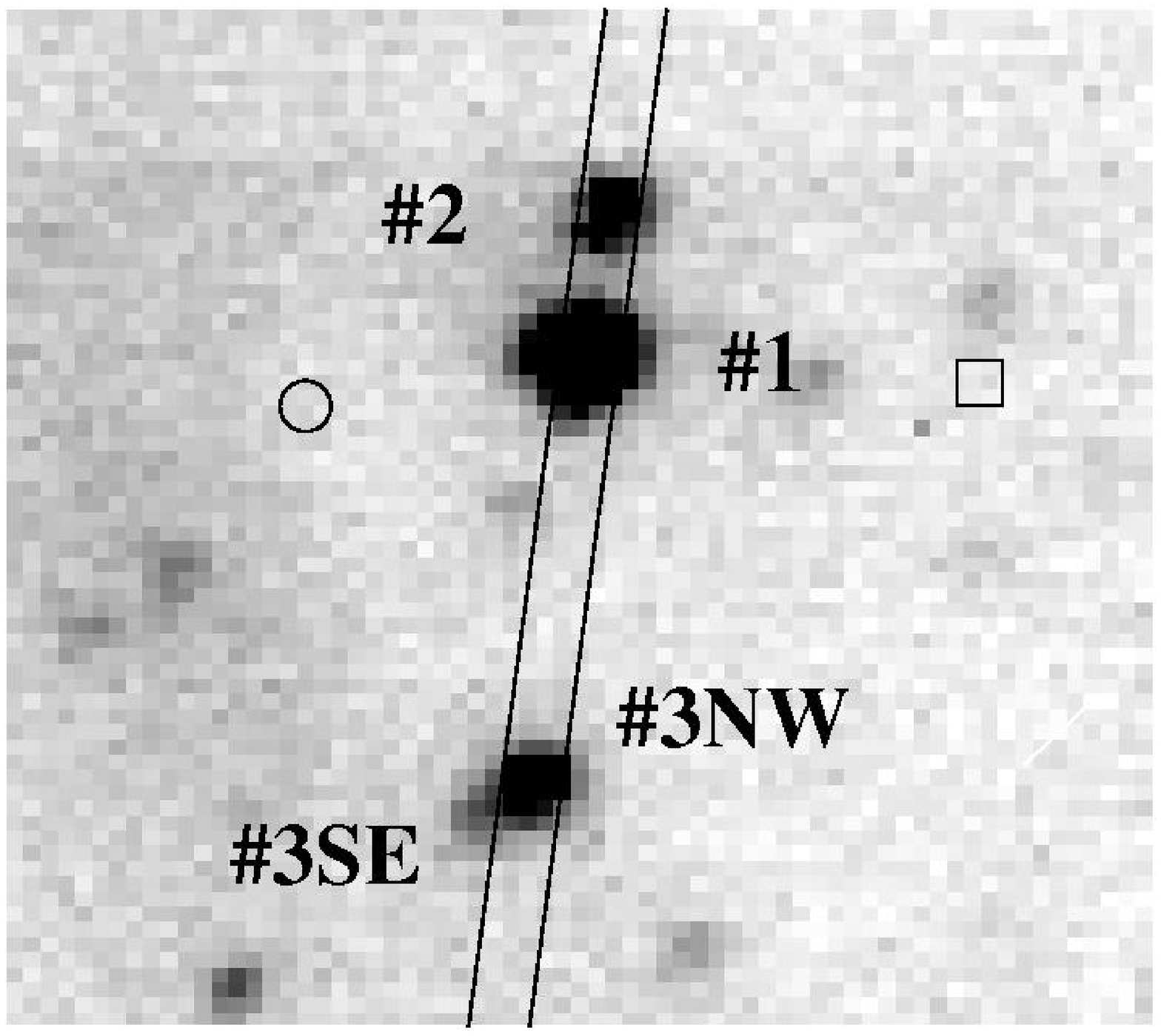}
      \caption{(left) 2$\times$2 arcmin colour composite JHK image centered
upon G23.96+0.15 from 2MASS; (right) 10$\times$10 arcsec  $\sim$2.18$\mu$m
acquisition image of G23.96+0.15 from ISAAC, in which the brightest
sources are marked  with the ISAAC slit overlaid. Source \#1 
has position $\alpha$ = 18:34:25.25, $\delta$ = --07:54:45.5 (J2000),
with the locations of the 6\,cm peak from Wood \& Churchwell (1989,
circle) and 21\,cm peak from Kim \& Koo (2001, square) also indicated. 
North is up and east is to the left.}
         \label{image}
   \end{figure}

In Fig.~\ref{image} we present a 2$\times$2 arcmin JHK composite image 
centered upon G23.96+0.15 from 2MASS plus the central 10$\times$10 arcsec 
of the $\sim$2.18$\mu$m ISAAC dataset. The ISAAC image reveals a bright 
source (\#1), located at 
$\alpha$ = 18:34:25.25, $\delta$ = --07:54:45.5 
(J2000.0). This source lies $\sim$2.7 arcsec west of the  6\,cm peak 
reported by Wood \& Churchwell (1989), though 3.5 acsec east of the
21\,cm peak of Kim \& Koo (2001) for this irregular UCHII region. Fainter 
near-IR sources lie 1.5 arcsec to the north of \#1 and 4.2 arcsec to the 
south, which we  shall refer to as sources \#2 and 3, 
respectively. Source \#3 is itself resolved into 
NW and SE components, separated by 0.7 arcsec. Photometry of these 
sources was  obtained from archival UKIRT UFTI images from 19 Jul 2000 
obtained with  the 1024$\times$1024 Hawaii array (0.09 arcsec/pix) and 
K-band (K98)  filter during seeing conditions of $\sim$0.6 arcsec, using a 
zero-point  obtained from 10 nearby 2MASS field stars, providing an 
accuracy of  $\pm$0.04 mag. The J-, H- and K- band magnitudes for 
the combined  sources \#1  and \#2 are 13.81, 11.46 and 9.94 mag, 
respectively (2MASS 18342523-0754455).


We have also inspected H-band images from the Steward 2.3m IRCam guider 
images of Hanson et al. (2002) using a 128$\times$128 Hawaii array (0.5 
arcsec/pix), obtained in seeing conditions of $\sim$1.5 arcsec, in which 
sources \#1 and \#2 are again blended, from which H(\#3) - H(\#1+\#2) 
$\sim$ 1.5~mag. Independent estimates of H(\#3NW)-H(\#2) $\sim 0.3$~mag 
and 
H(\#3NW)-H(\#1) $\sim 2.3$~mag have been obtained from our high spatial 
resolution ISAAC 1.71$\mu$m long-slit spectroscopy (see next 
section).  These sources were well aligned, so slit losses should 
minimal.  Photometric properties are presented in Table~\ref{tab1}, 
taking into account uncertainties resulting from  the ISAAC and IRCam 
observations.

\subsection{Near-IR spectroscopy}

Long-slit spectroscopy of G23.96+0.15 was obtained with ISAAC in April-May 
2006 at a position angle of 7.7 degrees west of north, in order to include 
sources \#1, \#2 and \#3NW at three medium resolution, 0.775\AA/pix) 
grating settings centered at 1.71, 2.09 and 2.20$\mu$m (recall 
Fig.~\ref{image}). These observations, obtained using a 0.6 arcsec 
slit width, were taken during  excellent seeing conditions (0.3--0.5 
arcsec) at low airmass (1.02-1.08) together with solar-type telluric 
analogues.

Six individual exposures, comprising three AB pairs, were obtained for each 
grating setting, with wavelength solutions achieved from comparison XeAr arc
datasets. From these, the observations covered 1.671--1.751$\mu$m, 
2.029--2.155$\mu$m and  2.140--2.265$\mu$m at spectral resolutions of 3.8\AA, 
6.0\AA\ and 6.0\AA\ respectively, as measured from arc lines. 

Telluric correction was achieved from a single AB pair of spectroscopic 
datasets of early-G dwarfs observed at a similar airmass to G23.96+0.15,
corrected for their spectral features using high resolution observations of
the Sun which were adjusted to both the radial velocity and spectral
resolution of the template stars.

The 2.09$\mu$m setup suffered from low-level variable structure which was 
accentuated upon flat-fielding, so that only the two other settings were 
flat-fielded. Consequently, the continuum S/N achieved was $\sim$100 for 
the 1.71$\mu$m setting and 150 for the 2.20$\mu$m setting for \#1, but 
only 60 for the 2.09$\mu$m setting. The continuum S/N for sources \#2 and 
\#3NW was, at best, no greater than 40-50.

   \begin{figure*}
   \centering
   \includegraphics[width=\columnwidth,angle=-90]{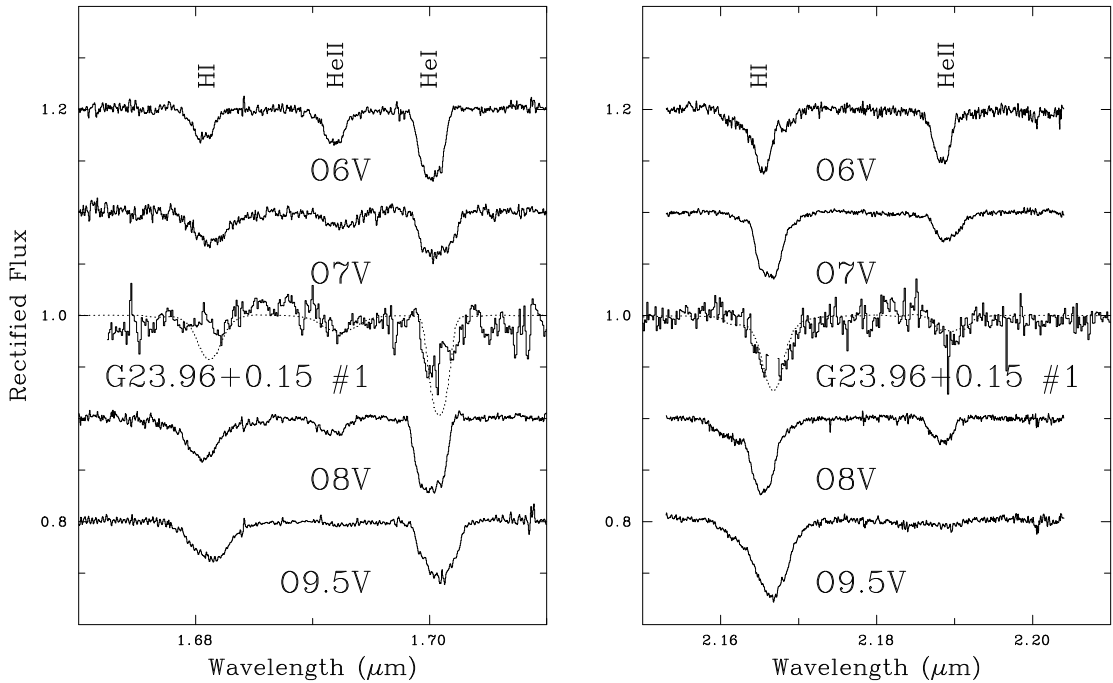}
      \caption{H-band (left) and K-band (right) VLT/ISAAC spectroscopy of 
G23.96+0.15 \#1 together  with MK template O6--9.5V stars from Hanson et 
al. (2005b), together with a spectral fit to \#1 (dotted line, see text)}
         \label{g23p96}
   \end{figure*}

\begin{table}
\caption{Near-IR properties of G23.96+0.15 sources from UKIRT/UFTI 
(K-band) and Steward 2.3m/IRCam imaging (H-band) plus 
VLT/ISAAC acquisition images and spectroscopy, from which 
equivalenth widths ($W_{\lambda}$ in \AA) were also measured for \#1.}
\label{tab1}      
\centering                          
\begin{tabular}{c@{\hspace{2mm}}c@{\hspace{2mm}}c@{\hspace{2mm}}c@{\hspace{1.5mm}}c@{\hspace{1.5mm}}c@{\hspace{1.5mm}}c@{\hspace{1mm}}c}        
\hline\hline                 
Source & K & H--K & $W_{\lambda}$(He\,{\sc ii}) & $W_{\lambda}$(He\,{\sc 
i}) 
&$W_{\lambda}$(H\,{\sc i}) 
&$W_{\lambda}$(He\,{\sc ii}) & Spect.\\
       & mag & mag & 1.692$\mu$m & 1.700$\mu$m & 2.165$\mu$m & 
2.189$\mu$m & Type\\
\hline                        
\#1 & 10.43 &  1.2 & 0.20 & 1.24 & 2.95 & 0.94 & O7.5V \\
    & $\pm$0.04&$\pm$0.1 &$\pm$0.08 & $\pm$0.11 & $\pm$0.17 & $\pm$0.12 \\ 
\#2 & 12.00 & 1.7 & \\ 
    & $\pm$0.04&$\pm$0.2 \\
 \#3NW & 12.10 & 1.9 & \\ 
    & $\pm$0.04 & $\pm$0.3 \\
  \#3SE & 13.7 &  --0.2 & \\ 
    & $\pm$0.2 & $\pm$0.3 \\
\hline                                   
\end{tabular}
\end{table}

\section{The ionizing star of G23.96+0.15}

\subsection{Near-IR spectral classification}

Hanson et al. (1996) developed a classification scheme for O stars from low 
resolution K-band spectroscopy. Classification of early to mid-O stars was 
achieved from the presence of C\,{\sc iv} 2.08$\mu$m and N\,{\sc iii} 
2.11$\mu$m emission 
lines, although it proved to be problematic to distinguish between late O and 
early B stars. This technique was extended to the H-band by Hanson et al. 
(1998) who highlighted the diagnostic role of the hydrogen Br 11 1.681$\mu$m, 
He\,{\sc ii} 1.692$\mu$m and He\,{\sc i} 1.700$\mu$m lines.

In Fig.~\ref{g23p96} we present H-band and K-band spectroscopy of G23.96+0.15
\#1 together with MK classification O dwarfs from the high quality, 
medium resolution atlas of Hanson et al. 
(2005b). From visual inspection, a classification of O7--8 is estimated, in
agreement with the low resolution K-band spectrum of \#1 from Hanson et al. 
(2002). However, we have also investigated the possibility of using the
He\,{\sc ii} 1.692$\mu$m/He\,{\sc i} 1.700$\mu$m and He\,{\c ii}
2.189$\mu$m/Br$\gamma$ ratios
for quantitative classification, independent of the metal lines in the K-band
(see also Lenorzer et al. 2004; Repolust et al. 2005).

\begin{figure}
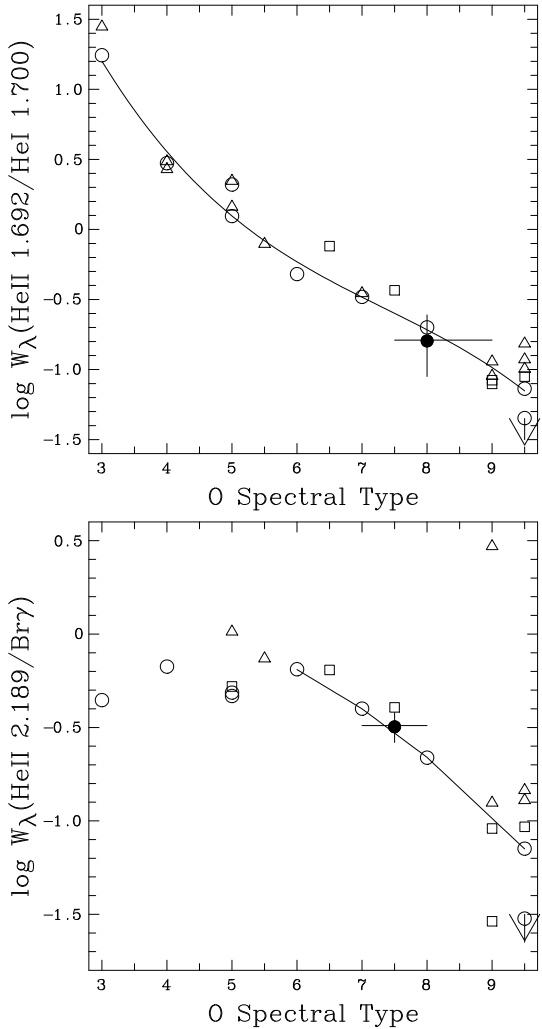

   \centering
   \includegraphics[width=0.8\columnwidth]{10942f3a.eps}
   \includegraphics[width=0.8\columnwidth]{10942f3b.eps}
      \caption{(Top) Equivalent width ratio of He\,{\sc ii} 
1.692$\mu$m/He\,{\sc i} 1.700$\mu$m
versus subtype  for O dwarfs (circles), giants (squares) and supergiants (triangles)
from Hanson et al. (2005a) and G23.96+0.15 \#1 (filled circle)
together with a polynomial fit to dwarfs as a guide (solid 
line); 
(bottom) as above for ratio of He\,{\sc ii} 2.189$\mu$m/Br$\gamma$}
         \label{hires}
   \end{figure}

We have measured the equivalent widths of these lines for MK classification O 
stars from Hanson et al. (2005b), whose absorption line ratios are presented 
in Fig.\ref{hires} (excluding Br$\gamma$ emission line supergiants). 
From this it is clear that the He\,{\sc ii} 1.692$\mu$m/He\,{\sc i}
1.700$\mu$m
ratio provides an excellent classification diagnostic for O stars, 
especially dwarfs, while the He\,{\sc ii} 2.189$\mu$m/Br$\gamma$ ratio 
provides
an additional constraint for late-type O dwarfs, providing the spectral 
resolution is sufficient to exclude any nebular contamination.  
Polynomial fits to equivalent width ratios of dwarf stars from Hanson et 
al. (2005b) are presented as a guide, excluding HD~37468 from the fit for 
which negligible He\,{\sc ii} absorption is observed. The
H-band ratio inherently represents a superior diagnostic, although the K-band 
ratio has two advantages for late-type O dwarfs, namely a lower interstellar
extinction plus 2.189$\mu$m is an intrinsically stronger He\,{\sc ii} line 
than 1.692$\mu$m.

Unfortunately, the moderate S/N of sources \#2 and \#3 prevented 
identification of spectral feaures beyond Br$\gamma$ nebular emission. We 
present H and K-band photometry of these sources in Table~\ref{tab1} 
together with absorption line measurements of G23.96+0.15 \#1. From the 
lower S/N H-band ratio we infer a dwarf spectral type of O8$^{+1}_{-0.5}$ 
while the K-band diagnostic indicates O7.5$\pm$0.5. Therefore, we propose 
an O7.5V classification for source \#1.

\subsection{Stellar temperature of G23.96+0.15 \#1}

The current spectroscopic calibration of O stars would result in 
an effective temperature of $T_{\rm eff}$=36\,kK for an O7.5 star (Martins 
et al. 2005). Instead, we have obtained a direct estimate of the stellar 
temperature  from a comparison between the near-IR hydrogen and helium 
lines and synthetic spectra obtained with the CMFGEN model atmosphere code 
(Hillier \& Miller 1998). 

CMFGEN solves the radiative transfer equation in the co-moving frame, 
under the additional constraint of statistical equilibrium. The 
temperature structure is determined by radiative equilibrium. Since CMFGEN 
does not  solve the momentum equation, a density or velocity 
structure is required.  For the supersonic part, the velocity is 
parameterized with a classical $\beta$-type law, with an exponent of 
$\beta$=1 adopted. This is connected to a hydrostatic density structure at 
depth, such that the velocity and velocity  gradient match at the 
interface. The subsonic velocity structure is set by a corresponding 
fully line-blanketed plane-parallel TLUSTY model (v.200, see Lanz \& 
Hubeny 2003). The atomic model is similar to that adopted in Hillier et 
al. (2003), including ions from H, He, C, N, O, Ne, Si, S, Ar, Ca and Fe.

We have assumed a depth-independent Doppler profile for all lines when
solving for the atmospheric structure in the co-moving frame, while in the 
final calculation of the emergent spectrum in the observer's frame, we 
have adopted a uniform turbulence of 50 km\,s$^{-1}$. Incoherent 
electron scattering and Stark broadening for hydrogen and helium lines
are adopted. Finally, we convolve our synthetic spectrum with a rotational 
broadening profile.

For an adopted terminal wind velocity of 2000 km\,s$^{-1}$, surface 
gravity of $\log g = 4$ and abundance ratio of He/H=0.1 by number, we 
varied the stellar radius and (non-clumped) mass-loss 
rate until a reasonable match to the He\,{\sc ii} 1.692$\mu$m, He\,{\sc i}
1.700$\mu$m, 
Br$\gamma$ and He\,{\sc ii} 2.189$\mu$m was achieved. An acceptable fit 
was 
achieved for $T_{\rm eff}$ = 38 $\pm$1\,kK, $\log \dot{M}/
(M_{\odot}$\,yr$^{-1}) = -6.3 \pm 0.2 $
and $v \sin i \sim 220 \pm 30$ km\,s$^{-1}$, as 
shown in Fig.~\ref{g23p96}. Nebular contamination is significant 
for He\,{\sc i} 1.700$\mu$m and high members of the hydrogen 
Brackett series. For an adopted mass of 30 $M_{\odot}$, 
a  radius of 9.2 $R_{\odot}$ results, which is representative of normal
Milky Way mid-O dwarfs (e.g. Repolust et al. 2005). We defer any further 
discussion of the physical properties of G23.96+0.15 \#1 until the next 
section where its distance is considered.



\begin{table}
\caption{Physical properties of G23.96+0.15 \#1
from our spectral analysis, based upon either the spectroscopic (top) or
kinematic (bottom) distances. Mass estimates are approximate since we 
adopt $\log g$=4.0. The predicted Lyman continuum ionizing flux ($Q_{0}$) 
is shown together with lower limits obtained from the 21\,cm radio 
observations of Kim \& Koo 
(2001,
$Q_{0}^{KK01}$) 
}
\label{tab2}      
\centering                          
\begin{tabular}{ccrcrccc}        
\hline\hline                 
$T_{\rm eff}$ & $\log L$ & $M$ & $\log \dot{M}$ &
$\log Q_{0}$ & $\log Q_{0}^{KK01}$ 
& $M_{\rm K}$ & $d$  \\
kK & $L_{\odot}$ & $M_{\odot}$ & $M_{\odot}$\,yr$^{-1}$ & 
s$^{-1}$ & s$^{-1}$ & mag & kpc \\
\hline                        
38 & 5.2 & 30 &$-$6.3 & 48.8 & $\geq$48.1 & --3.9 & 2.5  \\
38 & 5.8 &130 &$-$5.8 & 49.4 & $\geq$48.6 & --5.5 & 4.7 \\

\hline                                   
\end{tabular}
\end{table}

\section{Distance and extinction to G23.96+0.15}


Distances to Galactic H\,{\sc ii} regions are typically obtained from kinematic
methods, using an adopted rotation curve (Brand \& Blitz 1993)
and Solar galactocentric distance (8\,kpc, Reid 1993). Following this 
approach
the near (far) kinematic distance to G23.96+0.15 is 4.7\,kpc (10.0\,kpc) 
based upon the LSR velocity of 79.3 km\,s$^{-1}$ from Wink et al. (1983),
suggesting a distance of 4.2\,kpc from the Galactic Centre for 
G23.96+0.15. 
LSR velocities from NH$_{3}$ (Churchwell et al. 1990), CS (Plume et al. 1992) 
and H76$\alpha$ recombination lines observations (Kim \& Koo 2001) agree with
Wink et al. (1983) to within $\pm$1 km\,s$^{-1}$.


One would  expect  $M_{\rm K} = -3.9$ mag for G23.96+0.15 \#1 (O7.5V)
from the Conti et al. (2008) absolute visual magnitude-spectral type 
calibration, together with the intrinsic colours of Martins \& Plez 
(2006). We may 
correct its observed K-band magnitude for extinction using $A_{\rm K} = 
1.82^{+0.30}_{-0.23} E_{\rm H-K}$ (Indebetouw et al. 2005) and $(H-K)_{0} 
= -0.10$, from which $A_{\rm K} = 2.33 \pm 0.3$ mag, suggesting a 
distance modulus of 12.0$^{+0.3}_{-0.4}$, i.e. a distance of 
2.5$\pm0.4$\,kpc. For a representative scatter  of $\pm$0.5 mag in 
the absolute magnitude calibration of O stars, a  spectroscopic distance in 
the range 2--3.2\,kpc would be implied, placing G23.96+0.15 at a 
distance of 5.8$\pm0.5$\,kpc from the Galactic Centre. 

Mart\'{i}n-Hern\'{a}ndez et al. (2002) have previously estimated 
$A_{\rm K} = 2.0$ mag for G23.96+0.15 based upon HI 
recombination lines observed in Infrared Space Observatory (ISO) 
spectroscopy. We independently confirm this result from the observed ISO 
Br$\alpha$/Br$\beta$ ratio, although the ISO pointing was offset by 
$\geq$10 arcsec from the source peak, and so measured fluxes need to be 
treated with caution (Peeters et al. 2002).

In common with other studies of H\,{\sc ii} regions, we find a 
spectroscopic distance to G23.96+0.15 that is substantially lower than 
kinematic distances (e.g. Blum et al. 2001; Figueredo et al. 2008). The 
two distances could be reconciled if \#1 were substantially more luminous 
than typical mid-O stars, with $M_{\rm K} = -5.5$ mag, although this would 
require the star to be extremely massive if it is single. This is 
highlighted in  Table~\ref{tab2}, where physical properties of G23.96+0.15 
\#1 for each of  the alternate distances are presented. Of course, \#1 
could be an  unresolved equal mass binary (or compact star cluster). 
Such a source would appear $\sim$0.75 (0.3) mag brighter in the K-band than a 
single O7.5 star, owing to the contribution of its 
companion(s)\footnote{On the basis of a Salpeter-like Initial Mass 
Function for high mass stars}. A binary/cluster scenario would help 
to alleviate the spectroscopic and kinematic distance discrepancy, 
although in the  cluster case, stellar absorption lines from the O star 
would be diluted by the continuum light from lower mass cluster 
members, which does not appear to be the case for G23.96+0.15 \#1 
(recall Fig.~\ref{g23p96}).  Similar remarks apply for G29.96--0.02,
for which a spectroscopic distance of $\sim$3\,kpc would be expected 
for a single O4 dwarf using near-IR photometry from Pratap et al. (1999),
compared with a kinematic distance of 7.4\,kpc (Fish et al. 2003).

Kim \& Koo (2001) estimated an ionizing output of 10$^{48.8}$ ph\,s$^{-1}$ 
for G23.96+0.15A from their 21\,cm radio continuum observations, which is 
equivalent to an O7 dwarf (Conti et al. 2008), albeit based upon an 
assumed distance of 6\,kpc. If one was to adjust the distance to 4.7 
(2.5)\,kpc, a reduced Lyman continuum output would be obtained, 
representative of 
an O8 (O9.5) star (Table~\ref{tab2}). The measured 21\,cm flux, $\sim$0.75
dex lower than our indirect estimate,  is merely a lower limit since it may be
partially optically thick to bremstrahlung radiation and a fraction of 
the ionizing photons are likely to be absorbed by dust (either within the 
H\,{\sc ii} region or circumstellar cocoon). 

The near-IR colours of sources \#2 and \#3NW are consistent with being 
physically associated with G23.96+0.15 \#1 (\#3SE is likely a foreground
source). For the spectroscopic distance 
of 2.5\,kpc, they are probable early B dwarfs, at projected distances of 
0.02\,pc (\#2) and 0.05\,pc (\#3NW) from \#1. In common with indirect dust 
and  gas diagnostics for other UCHII regions (Okamoto et al. 2003), 
we find evidence that G23.96+0.15 likely hosts multiple massive 
stars (see also Hunter et al. 2004). For the kinematic (near) distance of 
4.7\,kpc, sources \#2 and \#3NW would possess absolute magnitudes typical 
of late O stars,  and so would also most likely be unresolved sub-clusters within the 
UCHII  region, at projected distances of 0.035--0.1 pc from \#1.

\section{Conclusions}

We show that G23.96+0.15 provides a second UCHII region whose ionizing 
star (\#1) is accessible to high S/N, medium resolution near-IR 
spectroscopy, complementing the higher ionization G29.96--0.02 source. We 
provide a quantitative near-infrared classification scheme for O stars, 
which has potential application for other visibly obscured O dwarfs in the 
Milky Way, from which a near-IR O7.5V classification is obtained for 
G23.96+0.15 \#1. We obtain a stellar temperature of $T_{\rm eff}$=38\,kK 
for 
G23.96+0.15 \#1 and infer a spectroscopic distance of 2.5\,kpc for a 
single star origin. This is substantially smaller than the (near) 
kinematic distance of 4.7\,kpc, in common with other obscured H\,{\sc ii} 
regions (e.g. Blum et al. 2001; Figueredo et al. 2008), although an 
unresolved binary or cluster nature would help to alleviate this 
discrepancy.  Two fainter sources are also likely early-type members of 
the UCHII region. The availability of spectral properties for G23.96+0.15 
should help empirically address the `inverse problem' of obtaining the 
ionizing O stars for other UCHII regions based solely upon mid-IR nebular 
characteristics.

\begin{acknowledgements}
Many thanks to Margaret Hanson for providing her H-band image of 
G23.96+0.15
and intermediate resolution near-infrared spectral atlas.  This 
publication makes use of data products  from 2MASS,  which is a joint 
project of the University of Massachusetts and the IPAC/CalTech, funded by 
the NASA and the NSF, and is, in part, based upon archival 
observations from UKIRT  which is operated by the Joint Astronomy Centre 
on behalf of the Science and Technology Facilities Council (STFC) of the 
UK. JPF acknowledges financial support from the STFC.

\end{acknowledgements}

\end{document}